# Bimodal directional propagation of wind-generated ocean surface waves

Paul A. Hwang[1*], David W. Wang[1], W. Erick Rogers[1]

Robert N. Swift[2], James Yungel[2] and William B. Krabill[3]

[1.]*Oceanography Division, Naval Research Laboratory, Stennis Space Center, MS 39529*

[2] *EG&G, N159, WFF, Wallops Island, VA*

[3] *NASA, WFF, Wallops Island, VA*

## Abstract

Over the years, the directional distribution functions of wind-generated wave field have been assumed to be unimodal. While details of various functional forms differ, these directional models suggest that waves of all spectral components propagate primarily in the wind direction. The beamwidth of the directional distribution is narrowest near the spectral peak frequency, and increases toward both higher and lower frequencies. Recent advances in global positioning, laser ranging and computer technologies have made it possible to acquire high-resolution 3D topography of ocean surface waves. Directional spectral analysis of the ocean surface topography clearly shows that in a young wave field, two dominant wave systems travel at oblique angles to the wind and the ocean surface display a crosshatched pattern. One possible mechanism generating this bimodal directional wave field is resonant propagation as suggested by Phillips resonance theory of wind wave generation. For a more mature wave field, wave components shorter than the peak wavelength also show bimodal directional distributions symmetric to the dominant wave direction. The latter bimodal directionality is produced by Hasselmann nonlinear wave-wave interaction mechanism. The implications of these directional observations on remote sensing (directional characteristics of ocean surface roughness) and air-sea interaction studies (directional properties of mass, momentum and energy transfers) are significant.

---

[*] *Corresponding author:* Dr. Paul A. Hwang, Oceanography Division, Naval Research Laboratory, Stennis Space Center, MS 39529. Email: phwang@nrlssc.navy.mil

# 1. Introduction

It is held as a common knowledge that wind generates waves, and that the wind-generated waves propagate primarily in the wind direction. Measuring the directional distribution of ocean waves, however, is a difficult task in ocean wave research. Based on the assumption that wind waves travel primarily in the wind direction, for the last forty years the directional distribution function of wind-generated waves is assumed to be unimodal. Field measurements using directional buoys or wave staff arrays all claim support of a unimodal directional distribution of a random wave field under steady wind forcing (e.g., Longuet-Higgins 1963; Mitsuyasu et al. 1975; Hasselmann et al. 1980; Donelan et al. 1985). While details of the results vary, the directional models established from these field investigations suggest that under steady wind forcing, waves of all length scales propagate mainly in the wind direction. The beamwidth of the directional distribution is narrowest near the spectral peak frequency and becomes broader toward both higher and lower frequencies. With such support of field measurements, unimodal directional distribution functions have been assumed in virtually all wave spectral models used in scientific and engineering applications.

The concept of unimodal directional distribution is now being questioned. Theoretical analysis of nonlinear wave-wave interaction results in a robust bimodal directional feature of the wave field (Banner and Young 1994). The nonlinear wave-wave interaction is the main mechanism responsible for frequency down shifting during wave growth. Wave energy at the spectral peak flows to lower spectral energy regions of higher and lower frequency components (e.g., Hasselmann et al. 1973). As nonlinear wave-wave interaction occurs among resonant quartet, spectral energy from dominant waves, which serve as the energy source of interaction, feeds to oblique components of the resonant quartet (Hwang and Wang 2001). This scenario is responsible for the robust bimodal directional distribution observed from numerical computations (Banner and Young 1994).

Verification of bimodal directionality can be obtained from analyzing temporal measurements using maximum likelihood method (MLM) or maximum entropy method (MEM). The quantitative description on the bimodal properties such as the lobe angle and lobe ratio (Banner and Young 1994), however, remains less certain because the results of bimodality derived from temporal data sources (directional buoys or wave gauge arrays) differ significantly depending on the method chosen in the analysis procedure (e.g., Young 1994; Young et al. 1995; Ewans 1998; Hwang et al. 2000b; Wang and Hwang 2001). A more unequivocal verification of bimodal directional distributions of wave components both above and below the spectral peak wavenumber is provided by directional spectral analysis of 3D surface wave topography of a quasi-steady wind-generated wave field, measured by a high-resolution airborne topographic mapper (ATM, an airborne scanning lidar system) (Hwang et al. 2000a-c).

Recently, the ATM system acquired a dataset of fetch-limited wave evolution under quasi-steady wind forcing in the Gulf of Mexico. The directional distribution of the wave field is quite complex and displays strong fetch dependence. For example, Plate 1 shows three strips of the photographic images taken from the aircraft during the experiment. Panel (a) is at the near shore region with the coastline visible in the picture. The surface geometry is dominated by linear undulations. The orientation of the wavy feature is crosswind rather than along wind. Panels (b) and (c) are taken at further offshore locations. The fetch in (c) is longer than that in (b) as deduced from the presence of breaking whitecaps in picture (c) that suggests a more developed wave field at that location. The directionality of the wave field in (b) is very confusing, and it is extremely difficult to determine the dominant direction of wave propagation from the picture. Panel (c) shows a striking crosshatched pattern suggesting the presence of two wave systems crossing each other at an oblique angle. The directional distribution of the crosshatched wave pattern is clearly bimodal. The bimodal feature of these young waves differs from the one produced by nonlinear wave-wave interaction described in the last paragraph. In the case of a young wave field, it is the dominant waves that are traveling oblique to wind. This result further casts doubts on the validity of the conventional belief that wind-generated waves propagate in the wind direction, an idea that has served as the basis of many results and assumptions concerning ocean wave directionality. Examples include the formulation of the directional distribution function and the wind input term of the energy conservation equation.

In this paper, we present the analysis of directional distribution of fetch-limited wind waves. The result suggests that the present formulation of wind-wave generation mechanism, especially its directional dependence, requires major revisions. Section 2 presents the environmental conditions prior to and during the field experiment. The evolution of the fetch-growing wave field can be quantified from the 3D topography of the wave surface. The bimodal directional propagation of the dominant waves is clearly demonstrated in the 3D topographic data. In the near shore region, the normal of the two wave fronts are almost perpendicular to the wind vector. As fetch increases, the angle between wind and wave directions decreases, and the wave components veer toward wind as they grow. A quantitative analysis of the spatial and temporal evolutions of the directional bimodality is given in Section 3. Section 4 presents discussions on the possible explanations of the observed bimodal propagation of dominant waves and several other issues such as the effect of strong modification of young waves by a prominent quasi-linear feature, possibly related to Langmuir circulation. Finally, a summary is given in Section 5.

**2. Field experiment**

Fig. 1 shows the location map of the ATM wave mapping during a frontal passage in the Gulf of Mexico on November 5, 1998. The aircraft flies a simple racetrack pattern with two long legs (~ 42 km each) in the wind direction. Offshore of the ATM flight track in the northeastern Gulf of Mexico, the National Data Buoy Center (NDBC) maintains two buoy stations (ID 42036 and 42039). The buoy locations are also shown in Fig. 1. Each station is equipped with a heave-pitch-roll sensor system. However, due to a sensor malfunction, only non-directional wave data are available during the period of airborne wave measurements. Buoy 42036 is moored approximately 75 km south of the ATM tracks at a water depth of about 50 m (28.51° N, 84.51° W). Buoy 42039 is moored approximately 125 km southwest of the ATM tracks at a water depth of 283 m (28.78° N 86.04° W). Wave data from both buoy stations are derived from a strapped-down accelerometer measuring buoy's heave acceleration over a 20-minute period each hour. Local wind speed and direction are acquired by a propeller-type wind anemometer mounted on the mast of the buoy at approximately 5 m above the design waterline of the buoy hull. Average wind speed and direction are computed from an 8-minute segment of wind data measured immediately after the wave data acquisition. The hourly wave and wind data and other meteorological and oceanographic parameters are relayed through the Geostationary Environmental Satellite (GOES) to NDBC for further processing and quality control. Details of the NDBC meteorological and wave measurement systems are described in Earle (1996).

In general, the wind and wave data from the two stations closely follow each other as shown in their time series (Fig. 2). In early November 4, the wind shifts to northerly after a frontal passage and starts to increase gradually. At midday of November 5, the northerly wind reaches 12 m s$^{-1}$ and the wave height is 2.1 m, which is the highest value during the period displayed in the figure. The ATM flight takes place between 1900-2120 UTC on November 5 when the wind speed remains almost constant at about 10 m s$^{-1}$ with a northwesterly direction. The sea states at both stations also remain quasi-steady. The wave height is about 1.8 m and the peak wave period 6 s at the buoy sites. The aircraft completes four repeat-track measurements with a time lag of ~0.47 h between consecutive repeat tracks. Thus the airborne measurement not only provides continuous spatial coverage but also temporal coverage at a moderate resolution. Both spatial and temporal evolutions of the wave development will be analyzed in the next section.

The sea state during this period is under the influence of a mild swell system generated by Hurricane Mitch. In mid-morning on November 4, Hurricane Mitch, weakened considerably from its peak stage, makes landfall over the northwestern Yucatan and later its center re-emerges in the south-central Gulf of Mexico (Fig.1). The storm regains strength and accelerates northeastward over the southeastern Gulf of Mexico. Along the path of the storm, long period swells of ~12 s are generated and propagate into the area of ATM site during the measurement period. The storm-generated swells are primarily from the southwesterly direction; the significant wave height of the swell system is approximately 0.3 m based on the ATM spectral analysis.

## 3. Bimodal propagation of dominant waves

*a. Spectral analysis of ocean wave topography*

As mentioned in the last section, the aircraft flies a repeat racetrack pattern with two long legs (~42 km each) aligned in the wind direction (Fig. 1). The flight track crosses a barrier island in the measurement site. In the following analysis, the fetch is measured from the ocean side of the barrier island coastline. Each continuous track (42-km coverage) is divided into segments of ~1.6 km length for directional spectral analysis. The swath of the surface topography is approximately 280 m. For this sequence of measurements, the circularly scanned data are resampled into square grids of 1.5 m on each side. The spectrum is calculated using a square area of 384m×384m of the surface topography (zeroes are filled in areas outside of the swath). Three independent spectra are derived from each 1.6-km segment (due to circular scanning, the leading and trailing edges of the segment are discarded, reducing the useful length of each topographic segment). Following the same procedure described in Walsh et al. (1985, 1989), the encounter wavenumber (due to that the measurement is taken from a moving platform) is corrected to true wavenumber by incorporating the information of flight velocity and local water depth. Also, smoothing is performed for each spectral element by averaging over a 3×3 region centered at the element. The weightings of 1, 1/2, and 1/4 respectively are applied to the central element, each of the four side elements, and each of the four corner elements. More details on the analysis procedure of ATM topographic data have been reported in Hwang et al. (2000a,c).

An example of the surface topography acquired by the airborne scanning lidar system near the offshore end of the flight track is shown in Fig. 3a. The surface relief depicts a crosshatched pattern produced by two wave trains crossing each other. The propagation directions of the two wave trains deviate significantly from the wind vector (shown as an arrow in the image). The directional wavenumber spectrum calculated from the 3D topography is shown in Fig. 3b. Bimodal propagation of the wave systems is clearly shown. The encounter wavenumber correction procedure produces the corrected wind wave spectrum in the right hand half-plane. For swell propagating against wind, the correct location of the swell component is on the left hand half-plane. For the example shown, the two spectral lobes on the right hand half-plane (marked with triangles) are corrected for Doppler wavenumber shift. The mirror images of the two lobes (marked with crosses) on the left hand half-plane are discarded in the subsequent spectral analysis. The corrected swell component, which travels against wind, is shown as a star on the left hand half-plane, and its image on the right hand half-plane is marked with a plus. The directional spectrum can be projected into the rectangular $(k, \theta)$ coordinates (Fig. 3c). When the plotting range of the abscissa is truncated, it is understood that the swell component shown is its image.

*b. Spatial and temporal evolutions*

Figure 4a shows a sequence of surface wave topography over the 42-km coverage. The corresponding directional wavenumber spectra for these image segments are shown on the right hand side of the figure (Fig. 4b). The directional distributions of this sequence of spectra show clear signatures of directional bimodality of dominant waves. In the near shore region, the dominant feature of the surface geometry is quasi-linear undulations with their directional normal perpendicular to the wind vector. Treating the two wave systems separately, the fetch distributions of the peak wavenumber and the peak angle of each system are shown in Fig. 5a and 5b. Several points are worth noting in this result. Firstly, over an extensive range of the fetch, the peak angle is close to 90°, that is, the dominant surface geometric feature is crosswind. Secondly, over the same fetch range, the peak wavenumber is within a very narrow range, basically between 0.2 to 0.3 rad m$^{-1}$. This wavenumber range is considerably lower than that expected from the fetch growth relations established from earlier measurements. For example, Young (1999) summarizes results of Hasselmann et al. (1973), Kahma and Calkoen (1992), Donelan (1992), and several other sources. He concludes that the field data can be fitted by the following envelop functions describing the dimensionless fetch dependence of wave energy and peak frequency

$$e_* = \max \begin{cases} (7.5 \pm 2.0) \times 10^{-7} x_*^{0.8} \\ (3.6 \pm 0.9) \times 10^{-3} \end{cases}, \quad (1)$$

$$f_* = \min \begin{cases} (2.0 \pm 0.3) x_*^{-0.25} \\ (0.13 \pm 0.02) \end{cases}, \quad (2)$$

where $e_*=\eta_{rms}^2 g^2/U_{10}^4$, $f_*=f_p U_{10}/g$, $x_*=gx/U_{10}^2$, $\eta_{rms}^2$ is the variance of surface elevation, $f_p$ is the peak frequency, $U_{10}$ is neutral wind speed at 10-m elevation, $x$ is fetch, and $g$ is the gravitational acceleration. Converting frequency to wavenumber using the linear dispersion relation, the mean and envelope described by (2) are superimposed in Fig. 5a. Over a large region from shore ($x$<~25 km) the measured peak wavenumber is much lower than that expected from (2). The result suggests that the crosswind feature observed may not be wind-generated progressive waves. It is possible that these longitudinal features aligned in the wind direction may be induced by Langmuir circulation. Further discussions of this aspect will be presented in the next section.

The directional distribution of the spectrum is calculated by

$$D(\theta,k) = \frac{\chi(k,\theta)}{\chi(k)}, \qquad (3)$$

where $\chi(k,\theta)$ is the directional wavenumber spectrum and $\chi(k)$ is the omnidirectional spectrum (Hwang et al. 2000a,b). To increase statistical stability, the directional distributions are further averaged over a 7-km distance (corresponding to 4 to 5 image segments except in the far and short fetch ranges where some of the tracks may have only two segments). As mentioned earlier, four repeat tracks are taken during the experiment. Reference to the time of the first track, the four tracks are referred to as measurements at time $t$=0, 0.47, 0.92 and 1.39 h hereafter. Fig. 6 shows the result at $t$=0.47 h. Since the left system is apparently modified significantly by the background swell, only the right system is displayed in the figure. The effect of the swell on wind waves will be discussed in a separate paper. In the nearshore segment (Fig. 6a, <$x$>=4.1 km), the dominant peak is cross wind. As fetch increases, the alignment of different wave components changes. The contour of the distribution develops an arched shape with higher wavenumber components curved toward the wind. The extent of the arch increases in the first half of the track. Using the 0.4 contour line as a reference, the arch extends from $0.55\pi$ to ~$0.08\pi$ at <$x$>=17.5 km (Fig. 6c). The arch feature then starts to recede. At the same time the dominant direction of the wave system moves into more windward direction (Figs. 6e and 6f). Fig. 7 shows the result at $t$=1.39 h. The development of the directional distribution is similar to that at the earlier time but the wave system is more organized and the directional spreading is considerably narrower, the wavenumber range defined by the 0.4-contour line is also smaller. At the last two fetch segments of this track, bimodal directional feature of nonlinear wave-wave interaction as observed for a quasi-equilibrium sea starts to develop, that is, wave components shorter than the peak wavelength display the bimodal feature of nonlinear wave-wave interaction (compare Fig. 7f with Fig. 3 of Hwang et al. 2000b).

The temporal evolution of the directional distribution at a given fetch can be better assessed from the time sequence presented in Figs. 8 and 9, showing the results at 10.2 and 38.7 km, respectively. At he shorter fetch (Fig. 8, $<x>$=10.2 km), the waves are generated over a broad directional range. As shown by the arched contours, the higher wavenumber components are more aligned with wind and lower wavenumber components are more toward the crosswind direction. As time progresses the directional range of the distribution decreases gradually. The wavenumber range over the four time steps remains similar. Based on the 0.4 contour line, the wavenumber range is from 0.2 to 0.65 rad m$^{-1}$. At the longer fetch (Fig. 9, $<x>$=38.7 km), the temporal development is similar to the case at the shorter fetch (Fig. 8) but both wavenumber and directional ranges of the contours are narrower. The directions of different wavenumber components are more aligned in the direction of the dominant peak, which is at an oblique angle from wind. At this longer fetch, the distribution shows the signature of directional bimodality due to nonlinear wave-wave interaction (bimodal distribution of wave components above the spectral peak).

To illustrate the spatial and temporal evolutions of the directional properties of wind-generated wave components during the development process, we trace the ridgelines of the 0.4 contour lines of the directional distributions. The results from all four repeat tracks are shown in Fig. 10. Quite interestingly, the temporal development of these ridgelines at any given fetch shows a high degree of similarity. At the shorter fetch ($<x>$=10.2 km), both directional and wavenumber ranges are very wide (0.15 to 0.55$\pi$, and 0.2 to 0.63 rad m$^{-1}$, respectively). The spectral energy of wind-generated waves is distributed along an arc on the ($k$, $\theta$) plane, shown by circles in Fig. 10. The data are least-square fitted with the following second order polynomial

$$\theta_* = -0.86129 k_*^2 - 0.18703 k_* + 0.61296, <x>=10.2 \text{ km}, \qquad (4)$$

where $k_*$ and $\theta_*$ are the $(k, \theta)$ coordinates of the ridgelines extending from the peak of the directional distribution function to the 0.4 contour line. At the middle fetch ($<x>$=24.5 km), the directional range remains almost the same but the wavenumber range shifts downward (0.1 to 0.46 rad m$^{-1}$). The second order polynomial fitting representing the ridgelines at this fetch is

$$\theta_* = -0.22791 k_*^2 - 0.94523 k_* + 0.59715, <x>\text{=24.5 km}. \tag{5}$$

At the longest fetch of the present experiment ($<x>$=38.7 km), the directional range of the ridgelines becomes narrower, mainly between $0.2\pi$ and $0.3\pi$, and the wavenumber range remains similar to the previous fetch. Bimodality of higher wavenumber components develops and two ridgelines can be traced. The second order polynomial fitting representing the ridgelines of the left and right lobes at this fetch is

$$\begin{array}{l}\theta_{*L} = -3.35875 k_*^2 + 1.22843 k_* + 0.17741 \\ \theta_{*R} = 0.53483 k_*^2 - 0.32908 k_* + 0.31879\end{array}, <x>\text{=38.7 km}. \tag{6}$$

The temporal and spatial evolutions of fetch-growth wave development under steady wind forcing are clearly much more complicated than the scenarios a unimodal directional distribution may have projected. From the airborne scanning lidar data, the analysis reveals that before reaching the equilibrium stage, the *dominant* waves are directionally bimodal. Under an ideal condition, two wave systems of approximately equal spectral energy propagate at large oblique angles with respect to the wind vector. The propagation angles decrease as fetch increases and the waves become more aligned in the wind direction. In addition to the directional bimodality of dominant waves, shorter waves are also generated in each wave system. At shorter fetches, the distribution of the generated short wave components of each wave system forms an arc on the $(k,\theta)$ plane. Combining the two systems (left and right), the wave spectral energy distributes more or less along a half circle on the $(k,\theta)$ plane (Figs. 8, 10). At larger fetches, the arch shaped distribution is still prominent. In addition, shorter wave components of each wave system develop bimodal distributions that show the character of nonlinear wave-wave interaction (Figs. 9, 10).

## 4. Discussions

### *a. The mechanism of directional bimodality of dominant waves*

The character of the bimodality of a young sea described in the last section is clearly different from the bimodality of a mature sea. The bimodal feature of the latter occurs at wave components shorter than the peak wavelength (e.g. Banner and Young, 1994; Young, 1994, Young et al. 1995; Ewans 1998; Hwang et al. 2000a, b; Wang and Hwang 2001). Hereafter we refer to the bimodal directional distribution of waves shorter than the peak component as the bimodal directional distribution of the first kind, and the bimodality of dominant waves as bimodal directional distribution of the second kind. These two kinds of directional bimodality are produced by very different mechanisms. The first kind directional bimodality is generated through nonlinear wave-wave interaction, in which wave energy near the spectral peak component feeds into both shorter and longer wave components with their wavenumbers and frequencies satisfying the condition of resonant quartet (Fig 11a) (Hwang and Wang 2001). In the present dataset, the directional bimodality manifests in the dominant components propagating at angles significantly different from the wind direction (Figs. 4b, 5b, 6-9). The most distinctive feature of the second kind directional bimodality is that the separation angles between the two dominate wave systems move closer to the dominant wind direction as the fetch increases. This is a signature of resonant propagation, in which wave components that maintain synchronized motion with the wind field receive continuous nourishment of energy and momentum from the wind and achieve more efficient growth. Because the phase speeds of waves at the earlier stage of wind wave generation are slower than the wind speed, to maintain at resonant propagation condition the waves travel at oblique angles to the wind vector (Phillips 1957). Phillips resonant generation theory of wind waves is originally developed for very short capillary-gravity waves at the initial stage of wind wave generation. The concept of resonant propagation to achieve efficient air-sea momentum transfer can be extended to much longer scale waves. The condition of resonance propagation can be expressed as

$$C_p = U_r \cos\theta_r, \tag{7}$$

where $C_p$ is the phase velocity of the dominant wave component, $U_r$ is the wind speed at a reference height proportional to the wavelength, and $\theta_r$ is the direction of resonant wave propagation with respect to the wind direction. For young waves, this resonance condition results in two wave systems propagating at oblique angles symmetric to the wind vector (Fig. 11b,c).

Evidence on the bimodal directional distribution from resonance wind wave generation is not very conclusive. Phillips (1958) suggests that the bimodal feature in the SWOP (Stereo Wave Observation Project, Cote et al. 1960) wavenumber spectrum is due to resonance wind wave generation mechanism. Similar conclusion is given by Jackson et al. (1985) on their directional wavenumber spectra obtained by an airborne imaging radar. In close inspection, the bimodal features observed in the above two datasets are in the wave components above the spectral peak wavenumber, and are similar to the observations presented in the airborne scanning lidar dataset (Hwang et al. 2000a,b). The generation mechanism of the bimodality is more likely due to nonlinear wave-wave interaction. A more convincing evidence would be to capture the wave generation at its young stage such that one can be observe the dominant waves propagating oblique to wind.

Walsh et al. (1985, 1989) present measurements from two wave-growth events using an airborne scanning radar system. The fetch coverage ranges from less than 100 to over 350 km. The nominal wind speeds are 17 and 12 m s$^{-1}$, respectively for the two datasets. The scanning radar yields continuous 3D surface topography of ocean waves along the flight track. The horizontal resolution (pixel size) of the system is 10 m with an absolute accuracy of ±0.3 m for the elevation measurement. The directional spectra are computed using segments of 10 and 22 km ground distances for the two experiments. One of the interesting results from these continuous measurements is that the peak of the spectrum is not in the wind direction until the fetch reaches approximately 150 km. Most of the spectra show only one peak in the directional distribution. They propose that the reason for the oblique angle between wave propagation and wind direction is the slant fetch effect resulting from complicated coastline geometry on the upwind region. Long et al. (1994) and Huang (1999) present a reanalysis of these directional spectra. The results show that the angular difference between the wind and wave directions is in agreement with the prediction of Phillips resonance wind wave generation mechanism. They also notice that some of the spectra do show two directional lobes, although in contrast to the theoretical prediction of two systems of equal energy, the two-lobe feature is rather asymmetric and in most cases absent. It is suggested that one of the two wave systems may have been suppressed by the ocean swell in the region.

Over the years, airborne ranging technology has advanced significantly. Major improvements include global positioning technology, the use of laser to reduce the footprint of the sensor, and the more compact and more powerful computers. The horizontal resolution of the ATM system is nonuniform; the coarsest is 6 m along track and 1.6 m cross track. The vertical resolution is 0.08 m (rms) (Krabill and Martin 1987; Krabill et al. 1995a,b; Hwang et al. 2000a,c). The ocean surface topography measured by the ATM measurements described in the last section show many characteristic features of resonant propagation. The most noticeable of them include: (a) In the nearshore region the wave fronts are almost perpendicular to the wind direction; as fetch increases waves grow longer and higher, the wave direction also changes from normal to wind to oblique to wind. (b) Since the resonance condition admits two solutions for the propagation angles, $\theta_r=\pm\cos^{-1}C_p/U_r$, two wave systems symmetric to the wind vector are developed. The surface topography displays a crosshatched pattern from the two systems crossing each other (Fig. 4a). While the ATM dataset show two wave systems, the spectral densities of the two are not the same. As suggested by Huang (1999) and Long et al. (1994), the background swell may be the cause of the asymmetry. The analysis of the swell influence on wind-generated waves measured by the ATM data is in progress.

Based on the ATM data, the calculated wind speeds satisfying the resonate propagation condition (7) are displayed in Fig. 12. The calculated wind speeds are in agreement with the buoy measurements of $U_{10}\approx10$ m s$^{-1}$ only in the far end of the flight track ($x>32$ km). Also, the partial agreement occurs only for the right-hand wave system, which is almost normal to the swell thus the swell modulation effect is expected to be minimal. For the whole track of the left system, which is more in the direction against the swell propagation, the measured angle between wind and the dominant wave component is much wider than expected from the resonant propagation condition. Possible reasons for the discrepancies are discussed next.

*b. Other environmental factors*

In Section 3, we notice that the dimensionless dominant wavenumber derived from the ATM dataset is much lower than that expected from the growth curve (2) established from extensive datasets of past research. An example is shown in Fig. 5a ($t$=0.92 h). In Fig. 13 we present $e_*(x_*)$ and $f_*(x_*)$ from all four flight tracks and compare the result with predictions of (1) and (2). As illustrated, in terms of the elevation variance, the agreement between the ATM data and calculation is very good (Fig. 13a). In the short fetch portion ($x_*$<1000), the ATM measured elevation variance is slightly higher than the envelop of the empirical relation. For $x_*$ between 1000 and 2000, the measurements are in excellent agreement with (1). This region is where many different experimental curves converge (e.g., see Figs. 5.8 and 5.9 of Young 1999). For the range $x_*$>2000, the measurements are between the mean and the lower envelop of (1).

The agreement in terms of the wavenumber or wave frequency does not fair as well (Fig. 13b). The dimensionless wave frequency analyzed from the ATM topography are obviously much lower than that predicted by (2) over the range of $x_*$<~2500, or $x$<~25 km. This range is also where major disagreement is found in the calculated and measured wind speeds satisfying the resonant propagation condition (Fig. 12b). The peak angles of the ATM spectra in this same region maintain close to 90 degrees (Fig. 5b). On the surface topography, these crosswind undulations manifest as quasi-linear features (Fig. 4a)

The characteristic length scale of the crosswind surface fluctuations is mostly between 15 to 40 m (calculated from $2\pi/k$). This length scale may suggest that the observed feature is likely related to Langmuir circulation pattern frequently observed in steady wind forced conditions. The connection between the Langmuir circulation pattern and nearshore wind-wave generation is not clear at this stage but the surface topographic measurements presented here strongly suggest that under steady wind forcing, linear undulations aligned in the wind direction represent a dominant feature of the ocean surface geometry at shorter fetches ($x<$ ~25 km for 10 m s$^{-1}$ wind speed).

*c. Implications on remote sensing and air-sea interaction*

The crosshatched pattern of surface waves is prevalent in a young wave field. The results analyzed in this paper raise a serious question regarding the directional properties of the surface roughness, which is a critical element in many remote sensing applications as well as studies of air-sea mass, momentum and energy transfers. For the former, scattering of electromagnetic waves for an active system and surface radiation for a passive device are sensitive to the orientation of the surface roughness, which is contributed by surface waves on the ocean surface. For air-sea interactions, directional properties of the momentum and energy fluxes between air and water is obviously dependent on the relative direction between winds and waves. Over the past four decades, spectral models assume unimodal directional distributions. Renditions of the ocean surface based on such directional models do not represent the ocean surface that is shown to be directionally bimodal. For example, field measurements repeatedly report a large crosswind to upwind ratio ($R$) of the mean square slope components (e.g., Cox and Munk 1954; Hughes et al. 1977). Cox and Munk (1954) report 9 cases of field measurements with natural or manmade slicks suppressing shorter waves. The range of the ratio $R$ is from 0.75 to 1.03. Such large crosswind mean square slope component cannot be explained using unimodal directional spectral models. The calculated result of the upwind and crosswind mean square slope components using bimodal directional distributions (of waves shorter than the peak wavelength) yield substantial improvement in agreement with field measurements (Hwang and Wang 2001). Also, a large crosswind to upwind ratio is expected for the condition of bimodal directional distribution of dominant waves as reported in this paper. By definition, if waves travel at 45° from the wind, $R=1$. As shown in Fig. 4, waves propagating at a large angle from wind are common for a young sea.

For another example, close correlation between the dynamic roughness ($z_o$) and root mean square wave height ($\eta_{rms}$) has been suggested in the sixties (e.g., Kitaigorodskii and Volkov 1965). Recent field measurements provide further confirmation relating $z_o$ and $\eta_{rms}$ (e.g., Donelan 1990; Donelan et al. 1993;

Anctil and Donelan 1996). Furthermore, Anctil and Donelan (1993) show that in addition to $\eta_{rms}$, incorporating the rms slope in the analysis improves significantly the correlation between measurements and the proposed empirical relation of $z_o$ and wave parameters.

The results derived from this study have significant implications on the formulation of the source and sink functions also. The data presented here suggest that in future development, the dynamic roughness needs to be considered as a vector property. It is fair to say that using the present formulation of wind input term (with a $\cos\theta$ directional dependence), it is impossible to produce the observed wave pattern shown in Fig. 4a. As it becomes more evident that the dynamic roughness is contributed by surface waves in length scales of active wind wave generation (e.g., Anctil and Donelan, 1996), it is reasonable to consider that the roughness elements are also directional. In particular, with the wave pattern showing clear bimodal features as displayed in Fig. 4a, one expects that the corresponding surface wind stress is also directionally bimodal. Consequently, the dynamic roughness should be decomposed in a similar way that one decomposes the physical roughness, which constitutes of surface waves. Although we do not have simultaneous wind measurements to pursue this issue further, it is noted that several earlier measurements have reported that the primary axis of the 2D probability distribution function (PDF) of small scale ocean surface roughness and the wind stress vector deviate from the wind direction. The angle of the major axis of the roughness PDF or wind stress vector deviating from the wind vector correlates well with the angle between background swell and the wind direction (Hwang and Shemdin 1988; Geernaert et al. 1993). The observation can be explained by the asymmetric (with respect to the prevailing wind direction) distribution of the two wave systems producing vector roughness deviating from the wind direction. Finally, the concept of decomposing surface wind stress into individual components, one for each discernable wave system, is recently applied to a SWAN model simulation for the bimodal resonant propagation condition. An example of the simulations is shown in Fig. 14, the resulting directional distributions of the fetch-growth wave fields are in good agreement with those shown in Fig. 4b. More details of the numerical simulations are reported in Rogers et al. (2000).

**5. Summary**

It has been held as a common knowledge that waves propagate in the direction of wind. This concept has been incorporated into unimodal directional distribution functions of virtually all spectral wave models. The formulation of wind input function in equations governing the dynamics of ocean surface waves also assumes a $\cos\theta$ function with $\theta$ measured from the wind direction. Recent field measurements of 3D ocean surface topography, however, do not support such assumptions. For young waves under fetch-limited conditions, the dominant waves travel at two angles significantly differently from the wind direction. In fact, at very short fetches the dominant wave direction is crosswind rather than along-wind. As fetch increases, two distinct wave systems propagate obliquely to wind can be identified from the crosshatched pattern of the surface topography (Fig. 4a). These patterns match the characteristics of resonant propagation between winds and waves. The resonant propagation is a key component of Phillips resonance generation mechanism of wind wave. At present, the resonant generation seems to be the only mechanism that provides a plausible explanation of the observed off-wind propagation of surface waves at the young stage of wind wave generation. Although nonlinear wave-wave interaction also results in energy transfer to off-wind components, the time scale (or equivalently the length scale representative of the fetch) for nonlinear interaction to take effect is long and the mechanism is probably not effective in shorter fetches. The measurement of the directional distribution (Figs. 8-9) indicates that characteristics of nonlinear wave-wave interaction become prominent only at $x > \sim 32$ km for $U_{10}=10$ m s$^{-1}$.

At a more mature stage, the directional distribution of waves shorter than the peak wavelength is also bimodal, forming two lobes symmetric to the dominant waves. The mechanism responsible for this bimodal directional distribution is nonlinear wave-wave interaction (Fig. 11a). Numerical computations of nonlinear wave-wave interaction (Banner and Young 1994), spectral analysis of 3D ocean surface topography (Hwang et al. 2000b), and analysis of crosswind to upwind components of surface roughness (Hwang and Wang 2001) all confirm the bimodal feature resulting from nonlinear energy transfer from the peak portion of the wave spectrum to oblique components of both longer and shorter wavenumbers of the wave spectrum.

In this paper, we refer to the bimodal directional distribution due to nonlinear wave-wave interaction as the first kind, and the bimodality with resonant propagation characteristics as the second kind. It is noted that the first kind bimodality is a very robust feature, as illustrated in the nonlinear simulations of Banner and Young (1994) and the subsequent proliferation of reported observations either from conventional temporal measurements using directional buoys (e.g., Young 1994; Young et al. 1995; Ewans 1998; Wang and Hwang 2001) or spatial measurements from the ATM system (Hwang et al. 2000a-c). In contrast, the second kind bimodality is apparently quite fragile. Earlier results from airborne scanning radar system (Walsh et al. 1985, 1989; Long et al. 1994; Huang 1999) show only one of the two expected off-wind propagating dominant wave systems that satisfy the resonant propagation condition. Swell suppression is suggested to be the mechanism wiping out the other resonant wave system. The airborne scanning lidar data presented in this paper show that under steady wind forcing, two dominant wave systems propagating at off-wind angles. The two systems are not symmetric as expected from the resonant propagation condition. In the offshore region where the crosshatched pattern is prominent, the spectral analysis shows that the development of windward evolution is significantly hindered in the wave system more aligned to a mild swell system (propagating opposing to wind at an oblique angle). The analysis of the other system, which is almost perpendicular to the swell, reveals that the propagation angles in the offshore region are in agreement with the calculation based on the resonant propagation condition (Fig. 12b). In the nearshore region ($x<25$ km), the dominant feature of the ocean surface topography is linear undulations with their directional normal almost perpendicular to the wind vector. The length scale of the feature is longer than the expected wavelength of wind-generated surface waves at the corresponding fetch. It is possible that these features are related to Langmuir circulation. The influence of Langmuir circulation, if it can be confirmed, also distorts significantly the bimodality of the second kind.

In spite of difficulties in measurements and observations, it can be concluded that bimodal directional distribution is a fundamental attribute of wind-generated ocean surface waves. In the mature stage, the bimodality (of the first kind) is robust and occurs in wave components shorter than the peak wavelength. The mechanism responsible for the bimodal generation is nonlinear wave-wave interaction. At the young stage of wind-wave generation, the bimodality (of the second kind) occurs in the dominant wave component and manifests in crosshatched surface geometry produced by two wave systems crossing each other. The second kind bimodality is fragile. Field data available to date indicate that swell or current shears such as those produced by Langmuir circulation can easily induce asymmetry to the bimodal distribution function or even destroy one of the two systems forming the directional bimodality (of the second kind).

**Acknowledgments**

This work is supported by the Office of Naval Research (Naval Research Laboratory Program Element N62435, "Phase Resolved Nonlinear Transformation of Shoaling Waves." (NRL Contribution JA/7330--00-0056).

**References**

Anctil, F., and M. A. Donelan, 1996: Air-water momentum flux observations over shoaling waves. *J. Phys. Oceanogr.*, **26**, 1344-1353.

Banner, M. L., and I. R. Young, 1994: Modeling spectral dissipation in the evolution of wind waves. Part I: Assessment of existing model performance. *J. Phys. Oceanogr.*, **24**, 1550-1571.

Cote, L. J., and Coauthors, 1960: The directional spectrum of a wind generated sea as determined from data obtained by the Stereo Wave Observation Project. *Meteor. Papers, New York Univ.*, **2**, W. J. Pierson (ed.), 88 pp.

Cox, C. S., and W. Munk, 1954: Statistics of the sea surface derived from sun glitter. *J. Mar. Res.,* **13,** 198-227.

Donelan, M. A., 1990: Air-Sea interaction. *The Sea*, Vol. **9**, B. LeMehaute and D. M. Hanes, Eds. Wiley, 239-292.

Donelan, M. A., F. W. Dobson, S. D. Smith, and R. J. Anderson, 1993: On the dependence of sea surface roughness on wave development. *J. Phys. Oceanogr.*, **23**, 2143-2149.

Donelan, M. A., J. Hamilton, and W. H. Hui, 1985: Directional spectra of wind-generated waves. *Phil. Trans. Roy. Soc. Lond.*, **A315**, 509-562.

Earle, M. D., 1996: Nondirectional and directional wave data analysis procedures. *NDBC Technical Document* 96-01.

Geernaert, G. L., F. Hansen, and M. Courtney, 1993: Directional attributes of the near surface wind stress vector. *J. Geophys. Res.*, **98**, 16571-16582.

Hasselmann, D. E., M. Dunckel, and J. A. Ewing, 1980: Directional wave spectra observed during JONSWAP 1973. *J. Phys. Oceanogr.,* **10,** 1264-1280.

Hasselmann, K. and Coauthors, 1973: Measurements of wind wave growth and swell decay during the Joint North Sea Wave Project (JONSWAP). *Deutsch. Hydrogr. Z.*, **A8**, 95 pp.

Huang, N. E., 1999: A review of coastal wave modeling: the physical and mathematical problems. In *Advances in Coastal and Ocean Engineering,* Ed. P. L.-F. Liu, World Scientific, Singapore, **4**, 1-20.

Hughes, B. A., H. L. Grant, and R. W. Chappell, 1977: A fast response surface-wave slope meter and measured wind-wave moment. *Deep-Sea Res.,* **24,** 1211-1223.

Hwang, P. A., and O. H. Shemdin, 1988: The dependence of sea surface slope on atmospheric stability and swell conditions. *J. Geophys. Res.,* **93**, 13903-13912.

Hwang, P. A., and D. W. Wang, 2001: Directional distributions and mean square slopes in the equilibrium and saturation ranges of the wave spectrum. *J. Phys. Oceanogr*. (in press).

Hwang, P. A., D. W. Wang, E. J. Walsh, W. B. Krabill, and R. N. Swift, 2000a: Airborne measurements of the wavenumber spectra of ocean surface waves. Part 1. Spectral slope and dimensionless spectral coefficient. *J. Phys. Oceanogr.*, **30**, 2753-2767.

Hwang, P. A., D. W. Wang, E. J. Walsh, W. B. Krabill, and R. N. Swift, 2000b: Airborne measurements of the wavenumber spectra of ocean surface waves. Part 2. Directional distribution. *J. Phys.*

*Oceanogr.*, **30**, 2768-2787.

Hwang, P. A., W. B. Krabill, W. Wright, E. J. Walsh, and R. N. Swift, 2000c: Airborne scanning lidar measurements of ocean waves. *Rem. Sens. Env.*, **73**, 236-246.

Jackson, F. C., W. T. Walton, and C. Y. Peng, 1985: A comparison of in situ and airborne radar observations of ocean wave directionality. *J. Geophys. Res.*, **90**, 1005-1018.

Kahma, K. K., and C. J. Calkoen, 1992: Reconciling discrepancies in the observed growth of wind-generated waves. *J. Phys. Oceanogr.*, **22**, 1389-1405.

Kitaigorodskii, S. A., and Y. A. Volkov, 1965: On the roughness parameter of the sea Surface and the calculation of momentum flux in the near-surface layer of the atmosphere. *Izv. Atmos. Oceanic Phys.*, **1**, 973-988.

Krabill, W. B., and C. F. Martin, 1987: Aircraft positioning using global positioning system carrier phase data. *Navig.*, **34**, 1-21.

Krabill, W. B., R. H. Thomas, C. F. Martin, R. N. Swift, and E. B. Frederick, 1995a: Accuracy of airborne laser altimetry over the Greenland ice sheet. *Int. J. Remote Sens.,* **16***,* 1211-1222.

Krabill, W. B., R. H. Thomas, K. Jezek, C. Kuivinen, and S. Manizade, 1995b: Greenland ice sheet thickness changes measured by laser altimetry. *Geophys. Res. Lett.*, **22**, 2341-2344.

Long, S. R., N. E. Huang, E. Mollo-Christensen, F. C. Jackson, and G. L. Geernaert, 1994: Directional wind wave development. *Geophys. Res. Let.*, **21**, 2503-2506.

Longuet-Higgins, M. S., D. E. Cartwright, and N.D. Smith, 1963: Observations of the directional spectrum of sea waves using the motions of a floating buoy. *Ocean Wave Spectra,* Prentice Hall, Englewood Cliffs, N. J., 111-136.

Mitsuyasu, H., and Coauthors, 1975: Observation of the directional wave spectra of ocean waves using a cloverleaf buoy. *J. Phys. Oceanogr.,* **5,** 750-760.

Phillips, O. M., 1957: On the generation of waves by turbulent wind. *J. Fluid Mech.*, **2**, 417-445.

Phillips, O. M., 1958: On some properties of the spectrum of wind-generated ocean waves. *J. Mar. Res.*, **16**, 231-240.

Rogers, W. E., P. A. Hwang, J. M. Kaihatu, and D. W. Wang, 2000: Modeling bimodal wind-wave propagation resonance. *Proc. 6$^{th}$ Int. Workshop on Wave Hindcasting and Forecasting,* Meteorological Service of Canada, 290-297.

Walsh, E. J., D. W. Hancock, D. E. Hines, R. N. Swift, and J. F. Scott, 1985: Directional wave spectra measured with the surface contour radar. *J. Phys. Oceanogr.*, **15**, 566-592.

Walsh, E. J., D. W. Hancock, D. E. Hines, R. N. Swift, and J. F. Scott, 1989: An observation of the directional wave spectrum evolution from shoreline to fully developed. *J. Phys. Oceanogr.*, **19**, 670-690.

Wang, D. W., and P. A. Hwang, 2001: Transient evolution of the directional distribution of ocean waves. *J. Phys. Oceanogr.* (in press).

Young, I. R., 1994: On the measurement of directional wave spectra. *Appl. Ocean Res.,* **16,** 283-294.

Young, I. R., 1999: *Wind generated ocean waves.* Elsevier, Amsterdam, 288 pp.

Young, I. R., L. A. Verhagen, and M. L. Banner, 1995: A note on the bimodal directional spreading of fetch-limited wind waves. *J. Geophys. Res.,* **100,** 773-778.

**List of figures**

Plate 1. Three photographs of ocean surface waves taken along a linear flight track, the fetch increases from top (near shore with coastline visible) to bottom, which is near the far end of the flight track.

Fig. 1. (a) Flight track at the experimental site in the northeastern corner of the Gulf of Mexico. The locations of NDBC Buoys 42036 and 42039 are marked. Also shown is the path of the remnant of Hurricane Mitch which generated background swell into the flight site. (b) Enlarged map near the experimental site.

Fig. 2. Wind and wave data from NDBC Buoys 42036 (+) and 42039 (×). (a) Wind speed, (b) wind direction, (c) wave height, and (d) wave period. A short line segment in each panel indicates the duration of ATM wave mapping.

Fig. 3. An example of the ATM measured surface topography and the resulting directional spectra. (a) 3D surface topography of ocean waves at ~38 km fetch. The wind is blowing from right to left in the coordinate system shown. (b) The directional spectrum calculated from the 3D topography shown in (a). Triangles are the peaks of two wave systems in oblique angles with the wind vector. Crosses are their images. The swell is shown with a star, and its image is shown with a plus sign. (c) The directional spectrum presented in the rectangular $(k, \theta)$ coordinates. The symbols used are identical to those used in (b).

Fig. 4. (a) 3D surface topography of ocean waves along one of the four flight tracks at 6 different fetches (38.1, 31.5, 24.8, 18.2, 11.5 and 4.93 km from top to bottom). The wind is blowing from right to left in the coordinate system of each topographic image. (b) The corresponding 2D spectra calculated from the surface topographies shown in (a). The wind direction is at $\theta = 0°$. A background swell system, shown by an arrow in each spectral panel, propagates against the wind.

Fig. 5. (a) The fetch development of the peak wavenumbers of the two wave systems (◁: left system, ▷: right system) analyzed from the ATM surface topography. The smooth curves are calculated based on Eq. 2 (Young 1999). Empirically, it is found that the wavenumber calculated from the first moment of spectrum, $k_1$ (shown as circles), is in better agreement with (2). (b) The angles

between the wind direction and the two spectral peaks.

Fig. 6. Spatial evolution of the 2D directional distribution function. Each plot represents the average over a segment approximately 7 km along the flight track. The time of this repeat track is 0.47 h after the first track. The average fetch of each panel is (a) 4.1 km, (b) 10.8 km, (c) 17.5 km, (d) 24.3 km, (e) 31.0 km, and (f) 36.8 km.

Fig. 7. Same as Fig. 6 but the time of track is 1.39 h after the first track. The average fetch is (1) 5.0 km, 10.2 km, (c) 17.0 km, (d) 24.5 km, (e) 32.0 km, and (f) 38.7 km.

Fig. 8. Temporal evolution of the 2D directional distribution function at $<x>=10.2$ km. Each plot represents average over a segment approximately 7 km along the flight track. The time is referenced to the beginning of the first flight track. (a) $t=0$, (b) $t=0.47$ h, (c) $t=0.92$ h, and (d) $t=1.39$ h.

Fig. 9. Same as Fig. 8 except that the average fetch is 38.7 km.

Fig. 10. The spatial and temporal evolutions of the directional distribution and wind wave development as depicted from tracing the ridgelines from the peak to the 0.4 contour of the distribution function. Data shown are at $<x>=10.2$ km (o: raw data, ★: polynomial fitting), $<x>=24.5$ km (□: raw data, ◇: polynomial fitting), $<x>=38.7$ km (left lobe: +: raw data, ◁: polynomial fitting; right lobe: ×: raw data, ▷: polynomial fitting).

Fig. 11. Two mechanisms producing bimodal directional distributions. (a) Nonlinear wave-wave interaction that transfers spectral energy near the spectral peak to oblique components at both higher and lower wavenumber components. (b) and (c) Slow propagating dominant waves align in directions oblique to the wind vector in order to maintain resonant propagation. The fetch of (b) is longer than that of (c).

Fig. 12. (a) The fetch development of the angle between the wind direction and the spectral peak (◁: left system, ▷: right system) analyzed from the ATM surface topography. (b) The wind speed satisfying the resonant propagation condition as calculated from the angle between the wind direction and the wave spectral peak. The wind speed measured by the ocean buoys is approximately 10 m s$^{-1}$ (shown as a solid curve) during the experiment.

Fig. 13 Comparison of (a) dimensionless wave variance, and (b) dimensionless peak frequency, as a function of dimensionless fetch. The curves (mean and envelop, Eqs. 1-2) of Young (1999) are also shown. o: $t$=0, ×: $t$=0.47h, Δ: $t$=0.92 h, □: $t$=1.39 h.

Fig. 14. An example of the directional spectra simulated with vectrorized surface wind stress components that are in the direction of expected wave directions satisfying the resonant propagation condition.

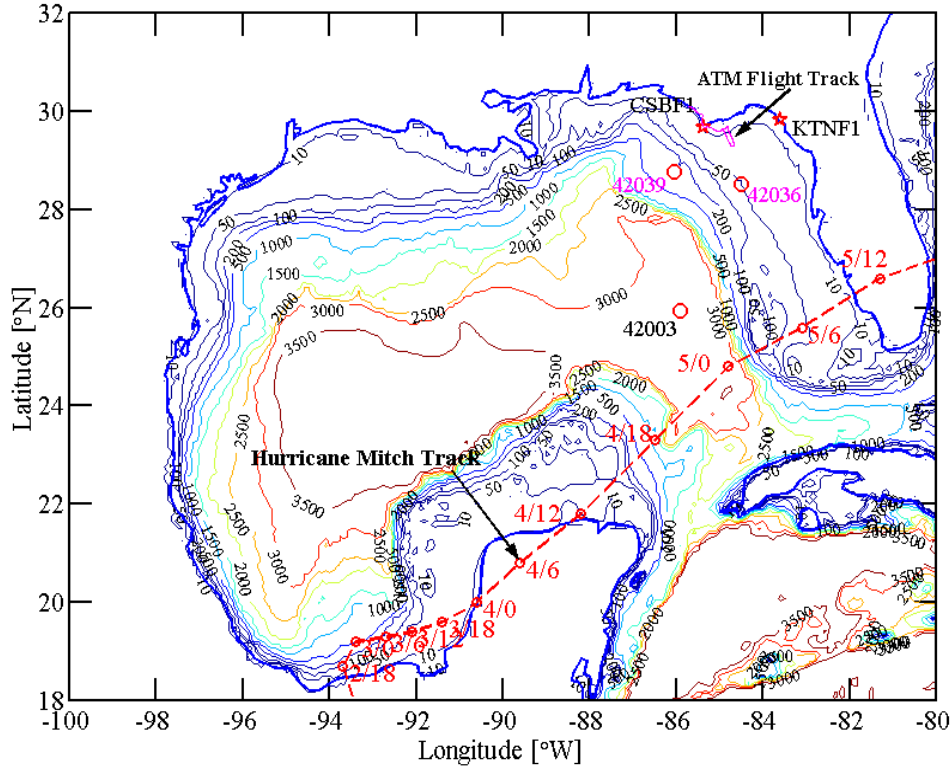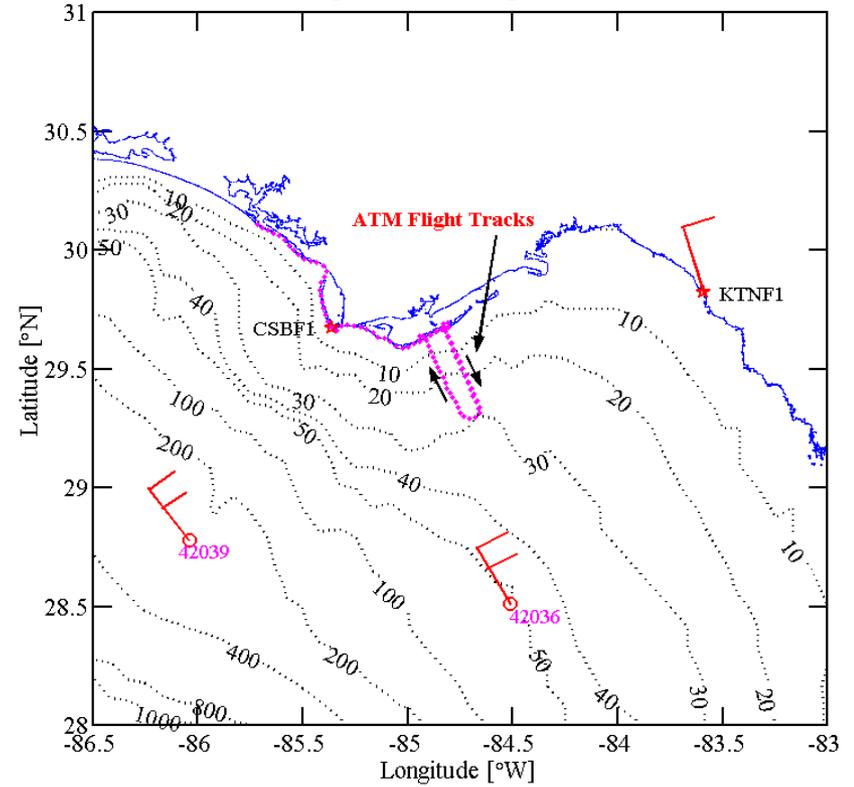

ATMGoM98Bimodal 1

w2buoygomx98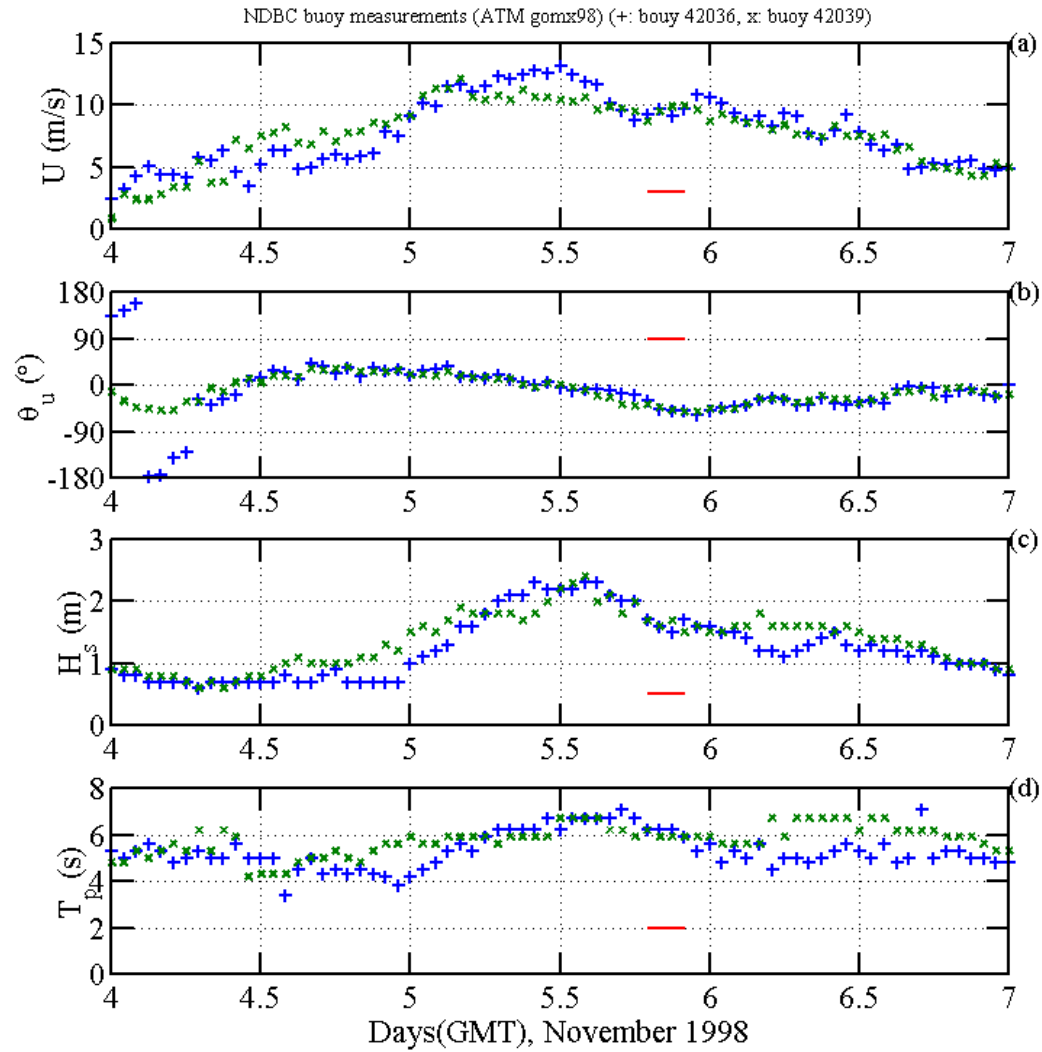

ATMGoM98Bimodal

Input: 200928, 201403, 25. d1,d2: 1.5, 1.5. time,lon,lat: 201404, 29.358, 275.237, 201430, 29.369, 275.228 981105va-61.4dep26.6

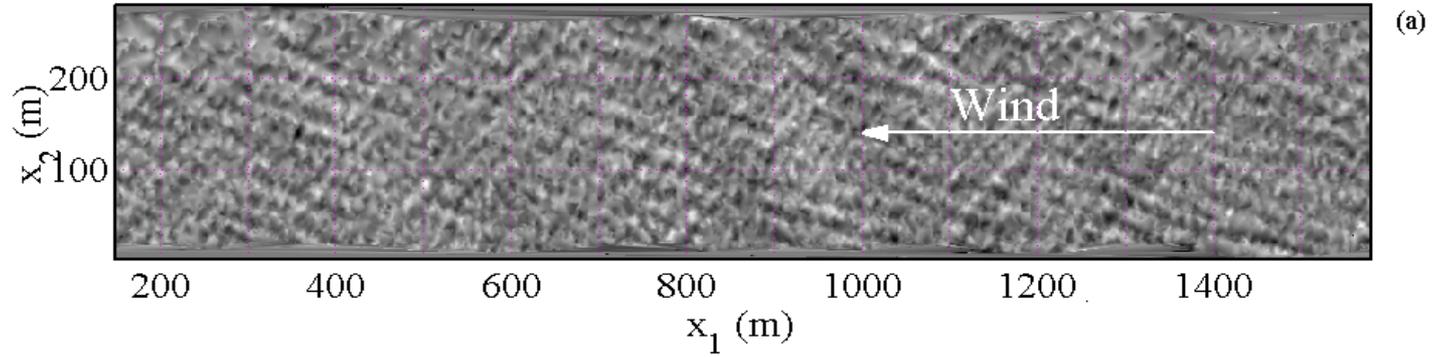

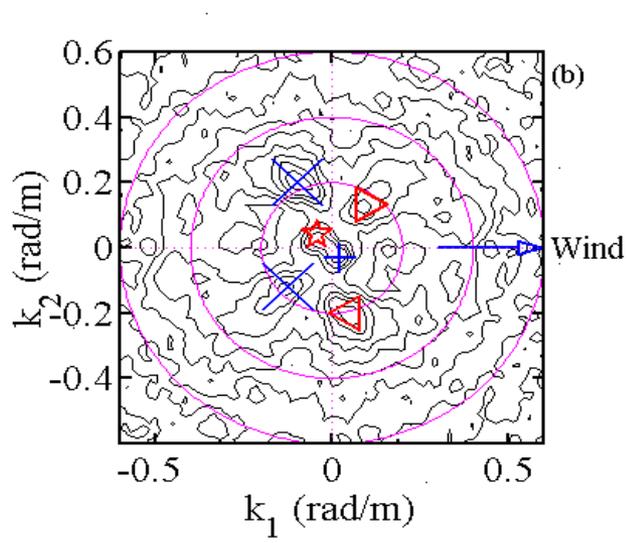
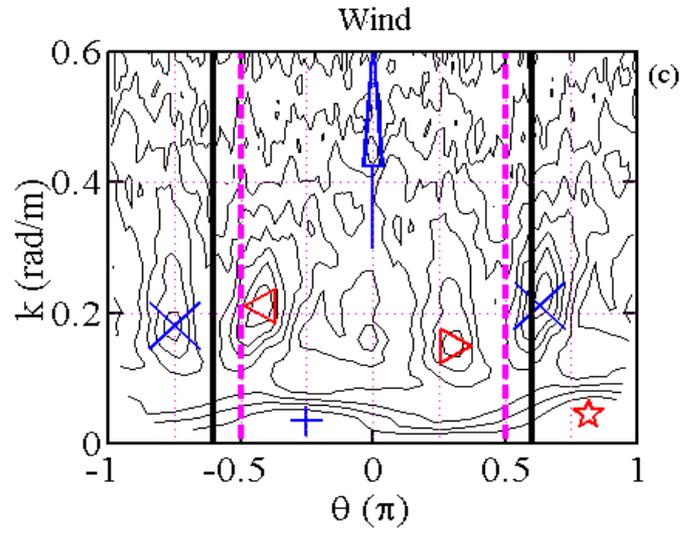

ATMGoM98Bimodal

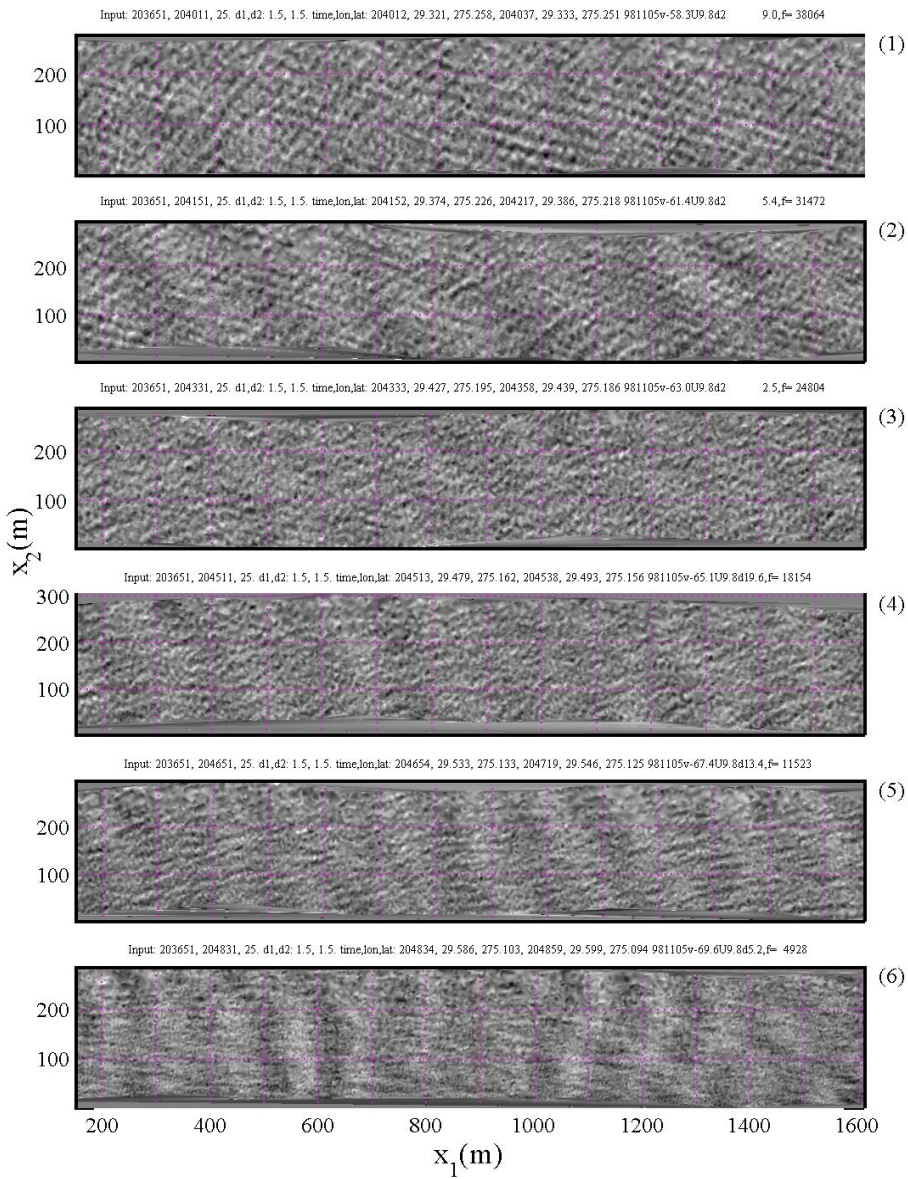
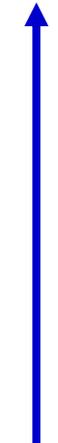

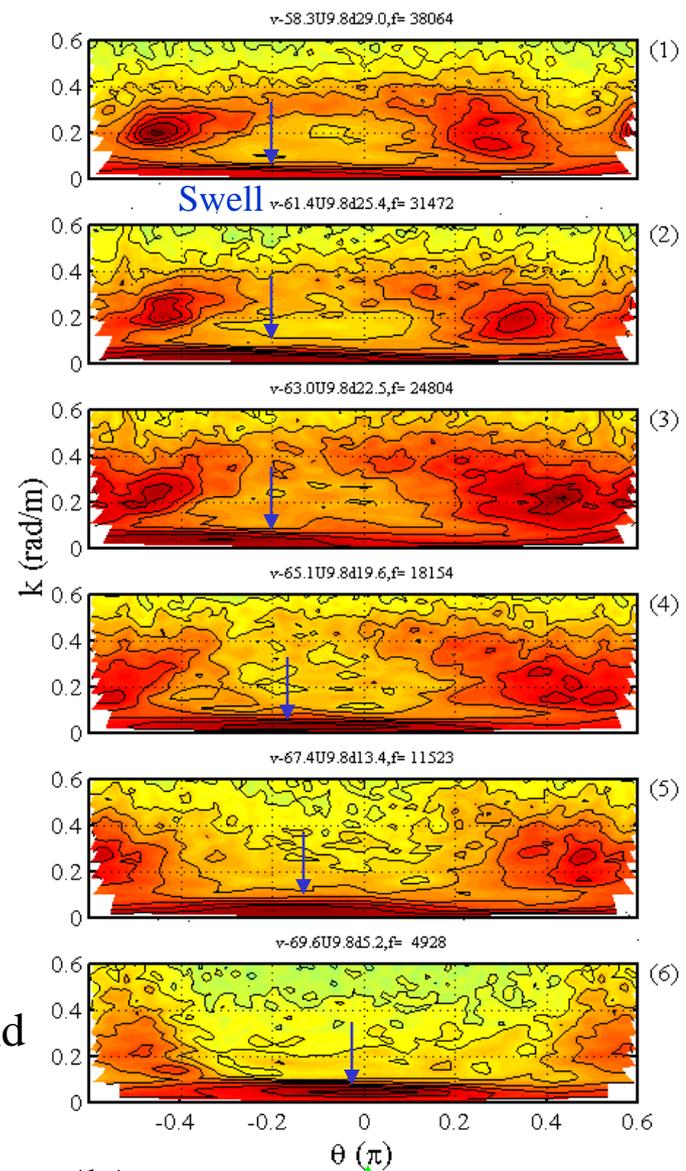
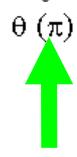

ATMGoM98Bimodal

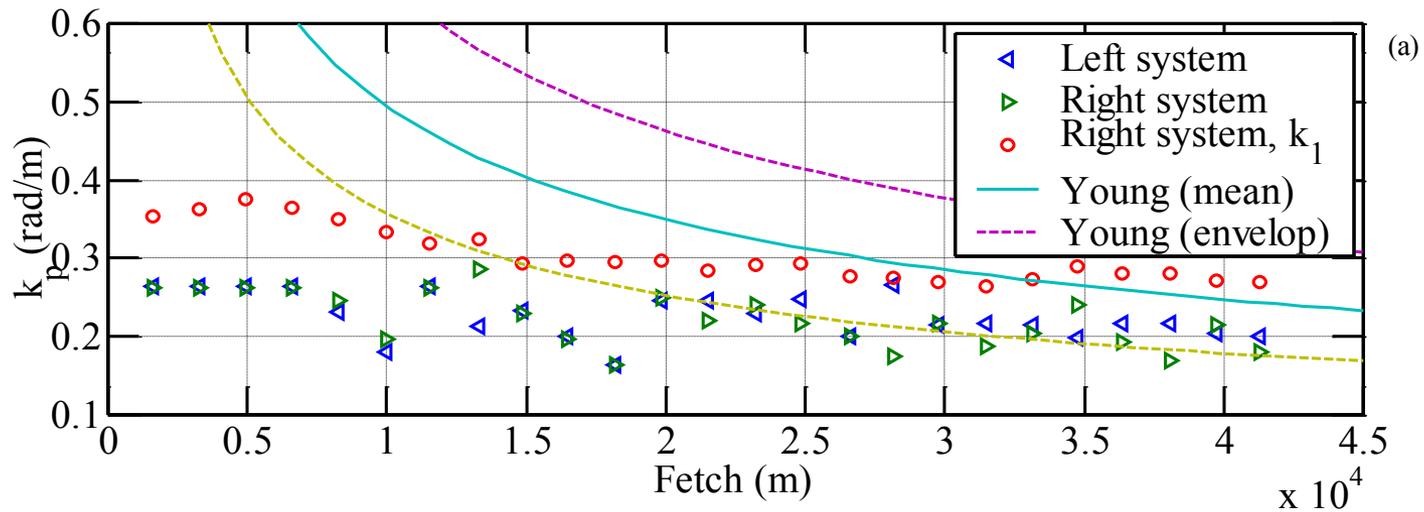
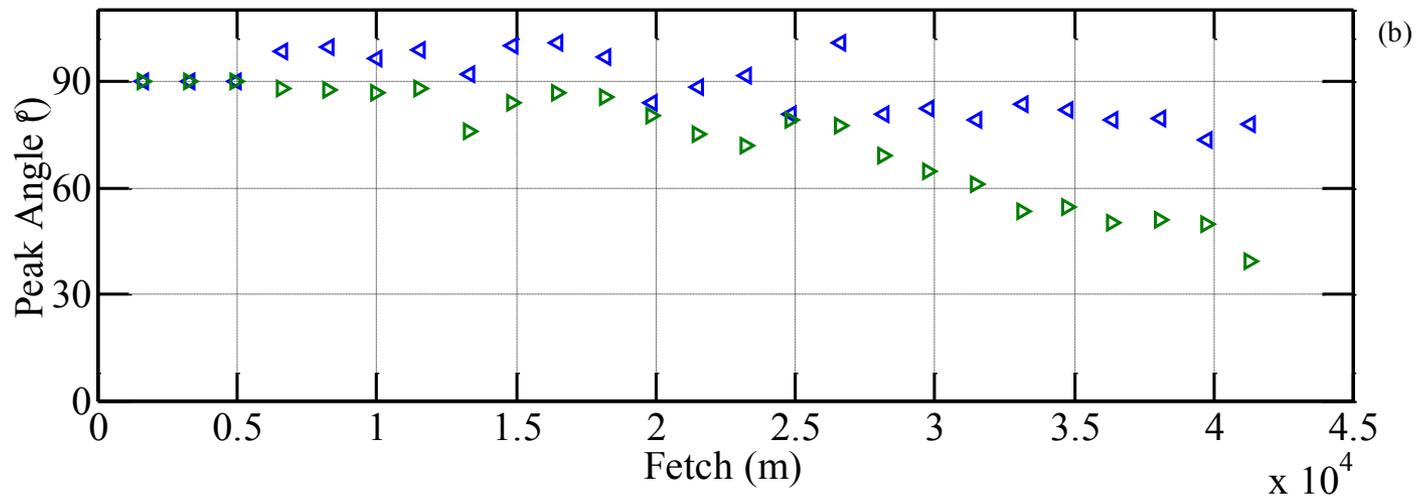

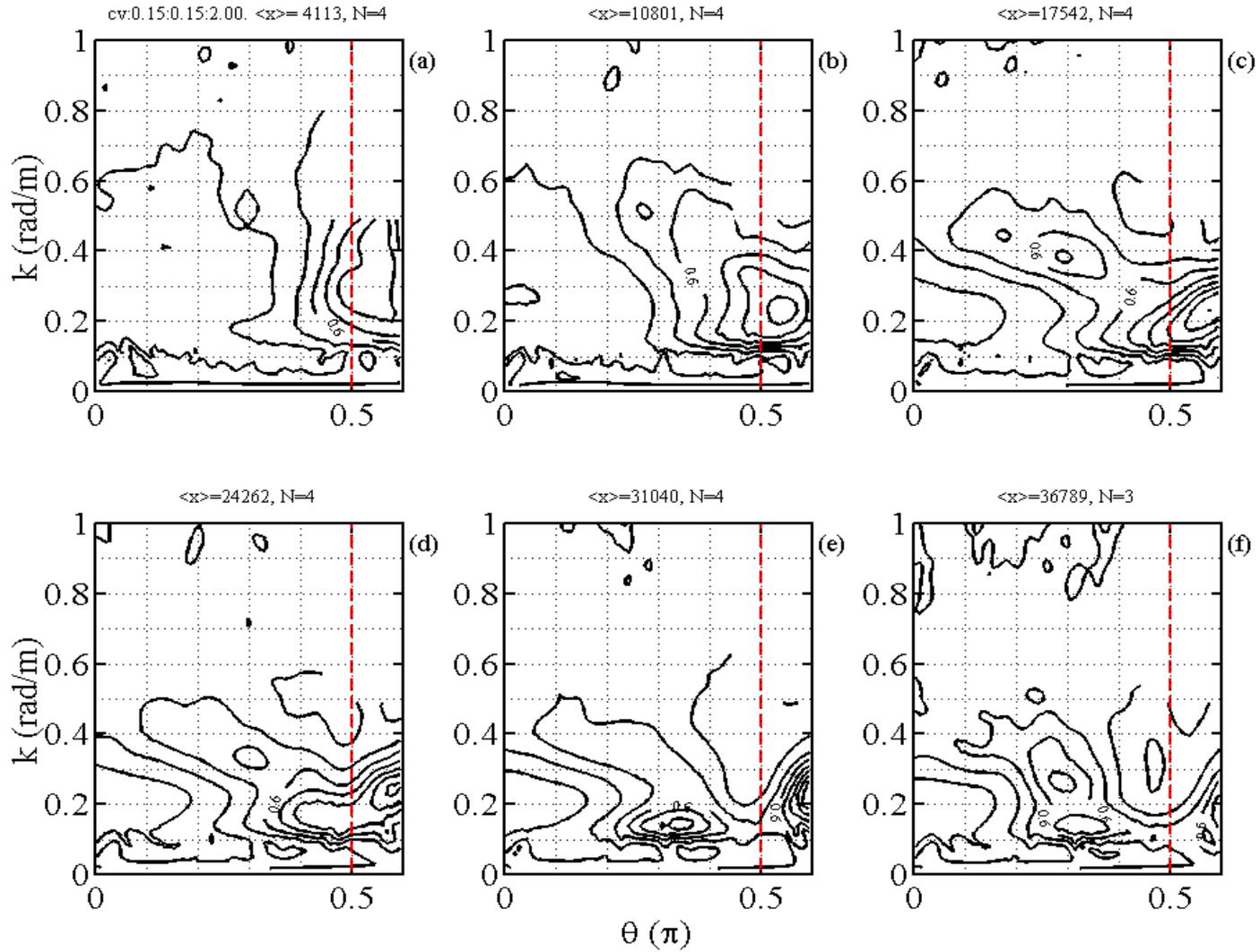

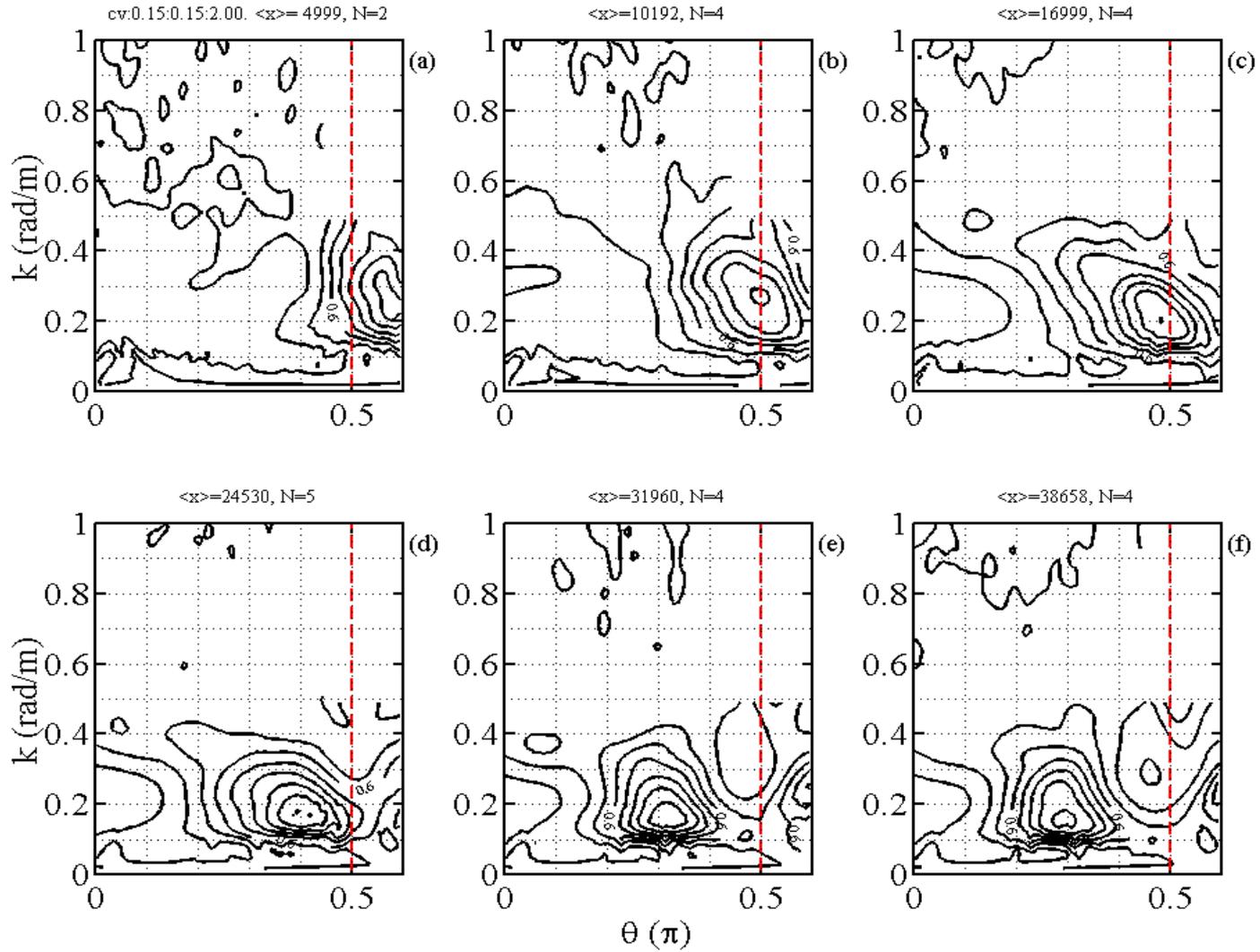

ATMGoM98Bimodal 7

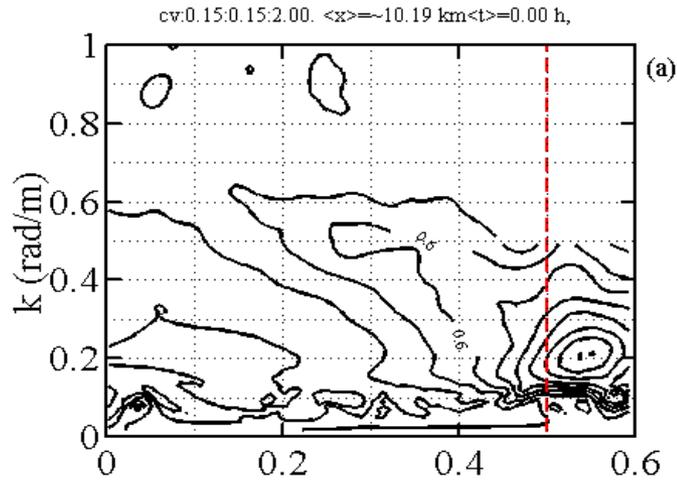 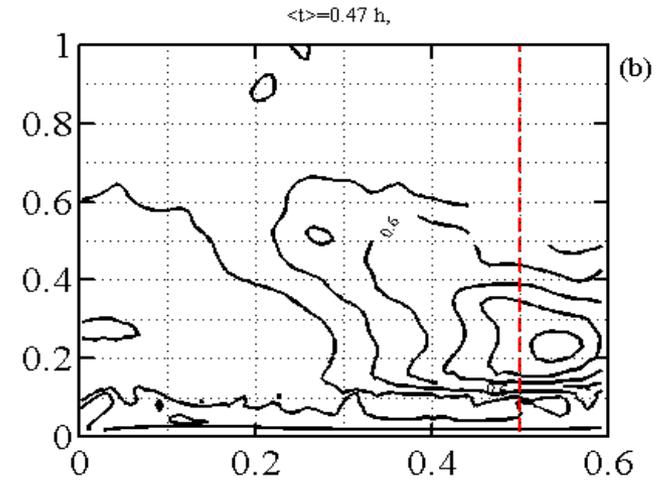
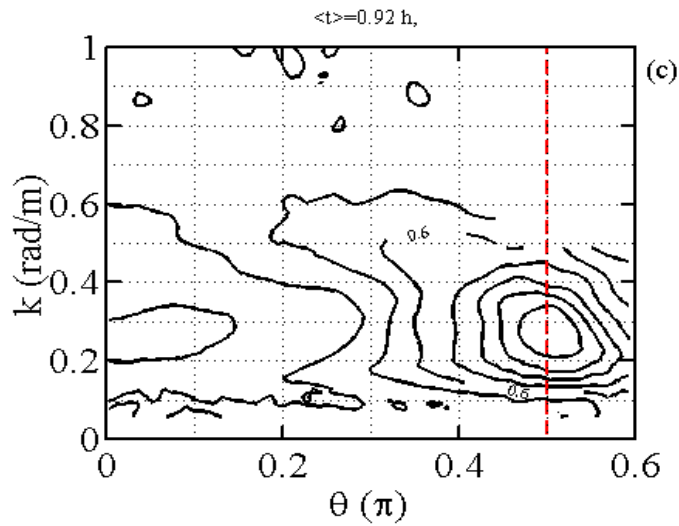 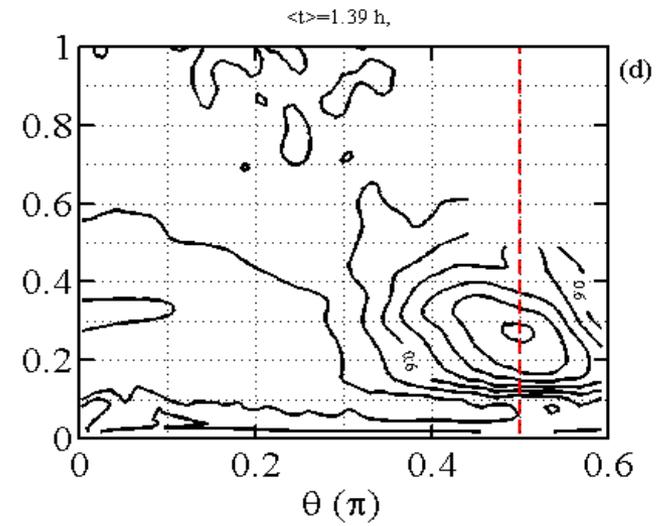

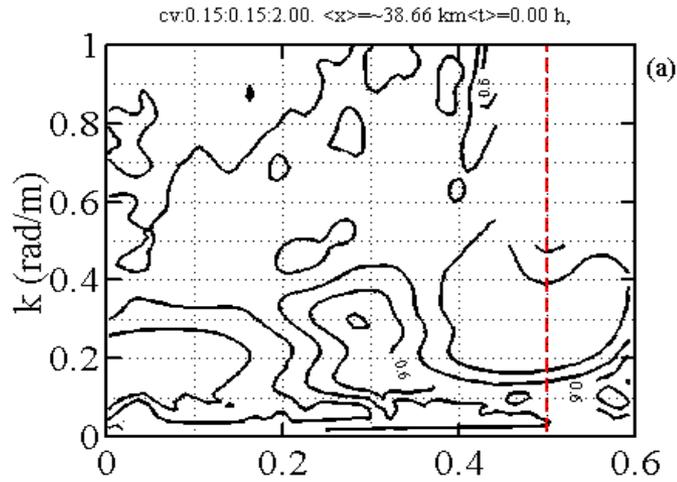
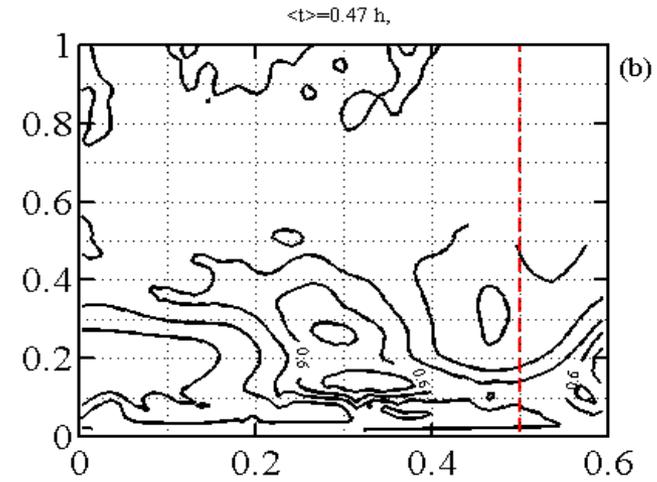
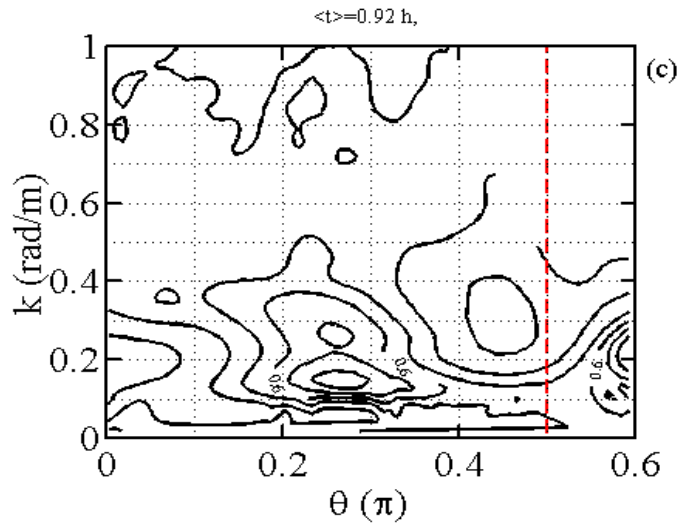
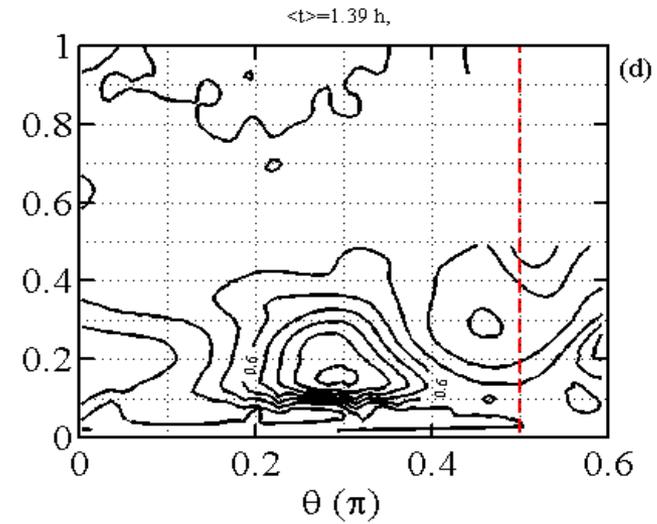

ATMGoM98Bimodal 9

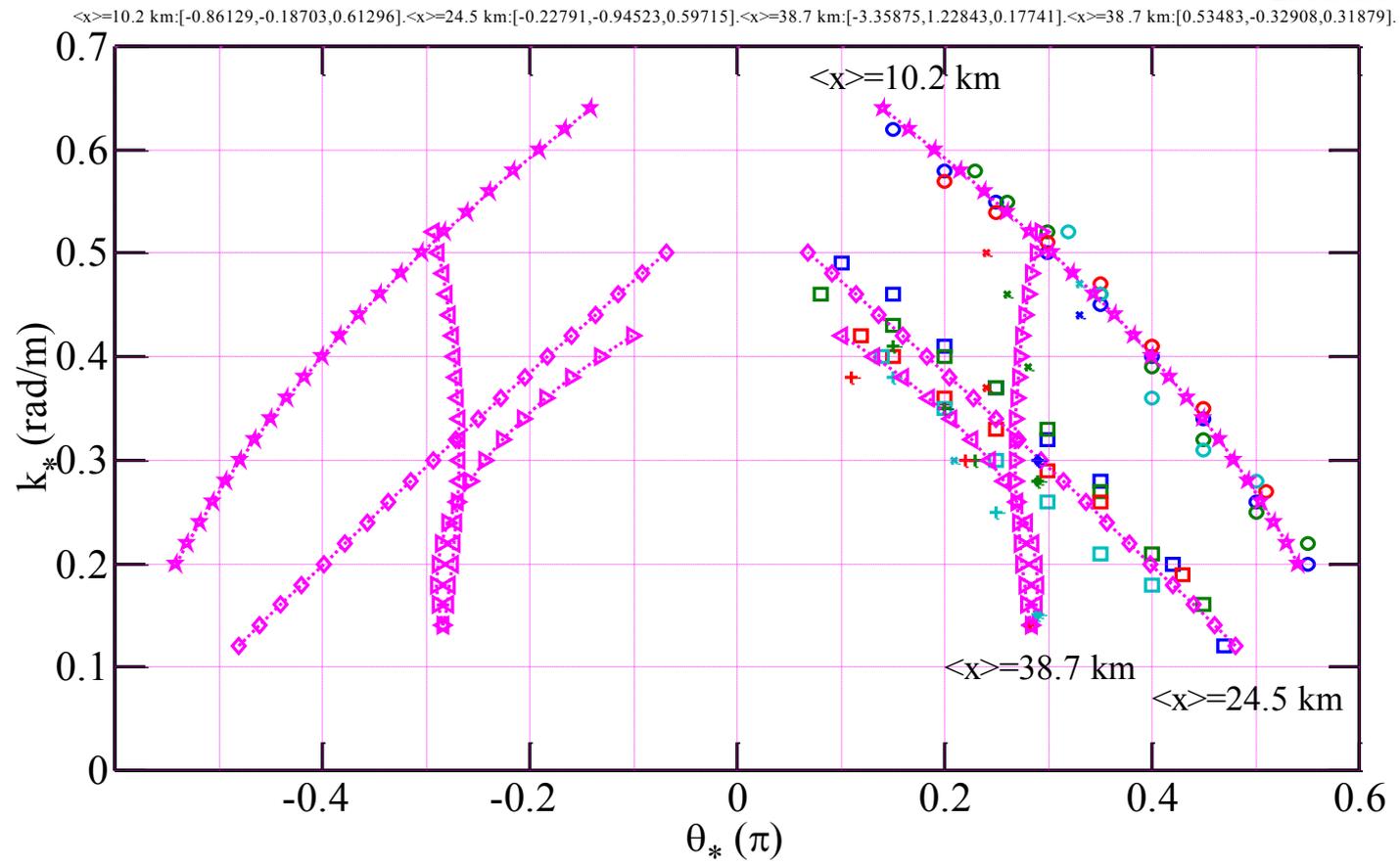

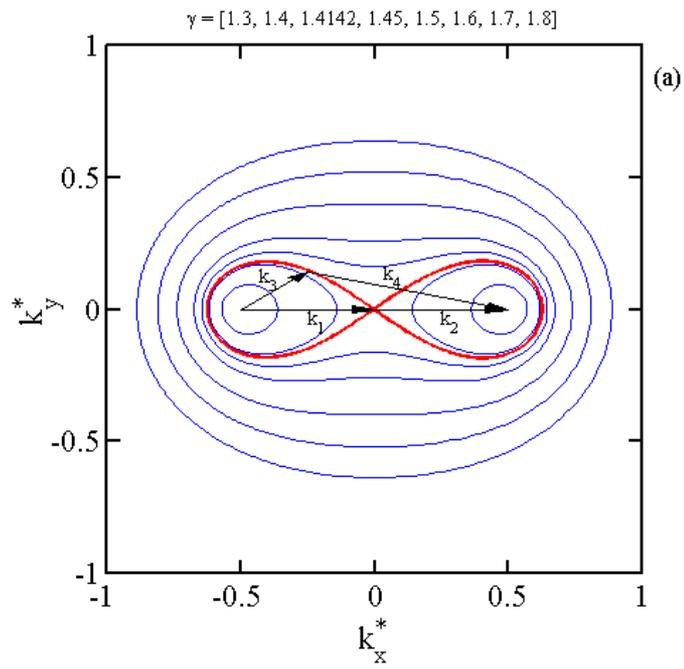
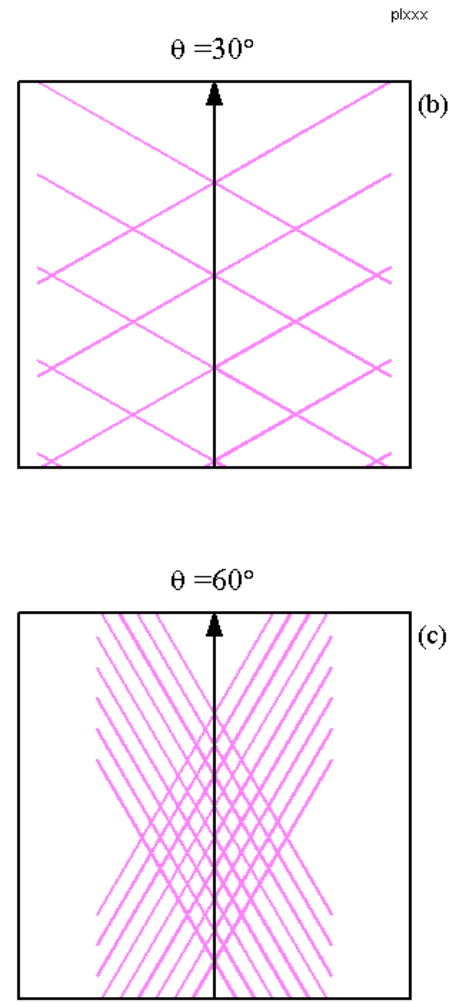

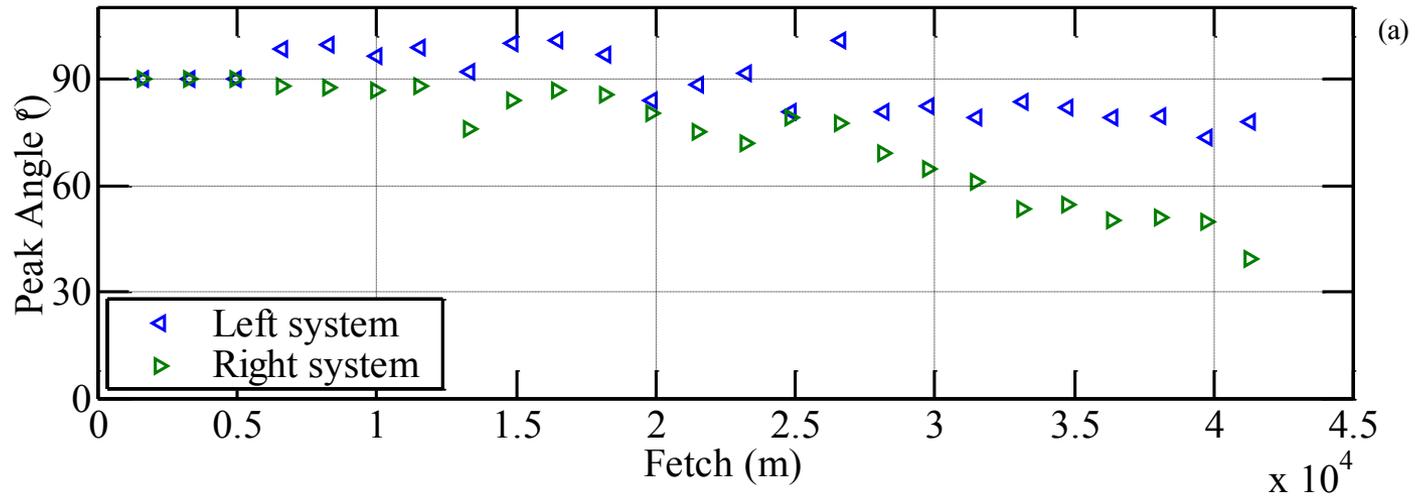
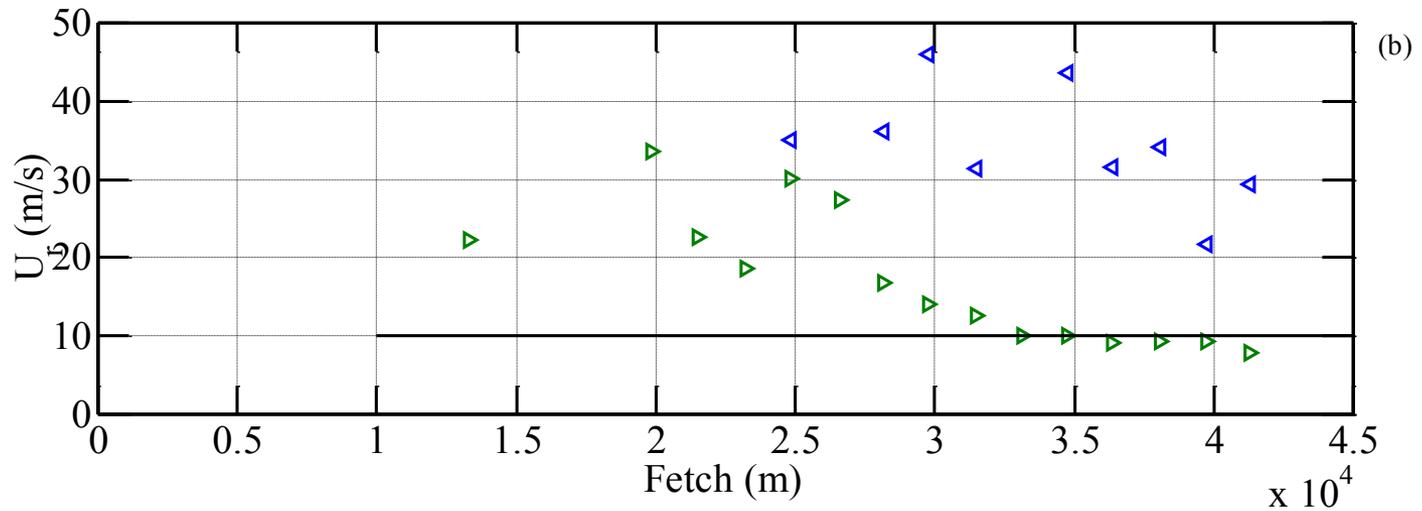

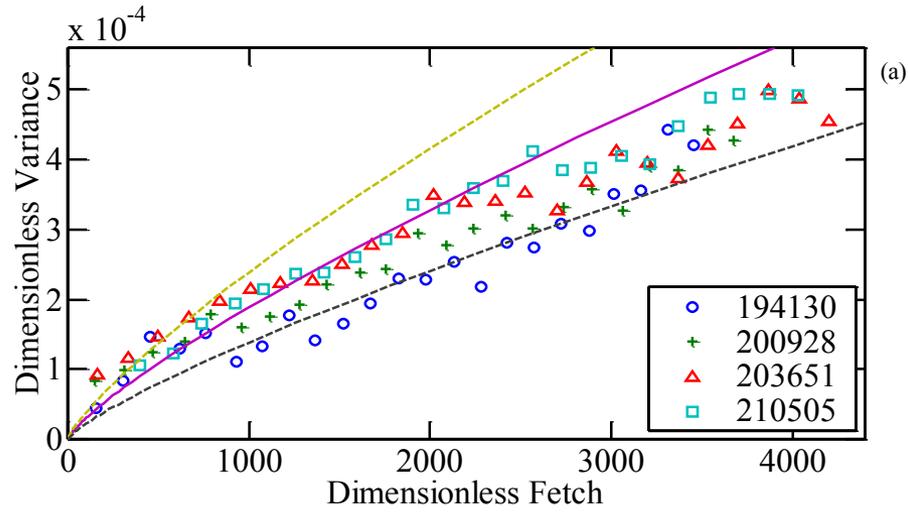

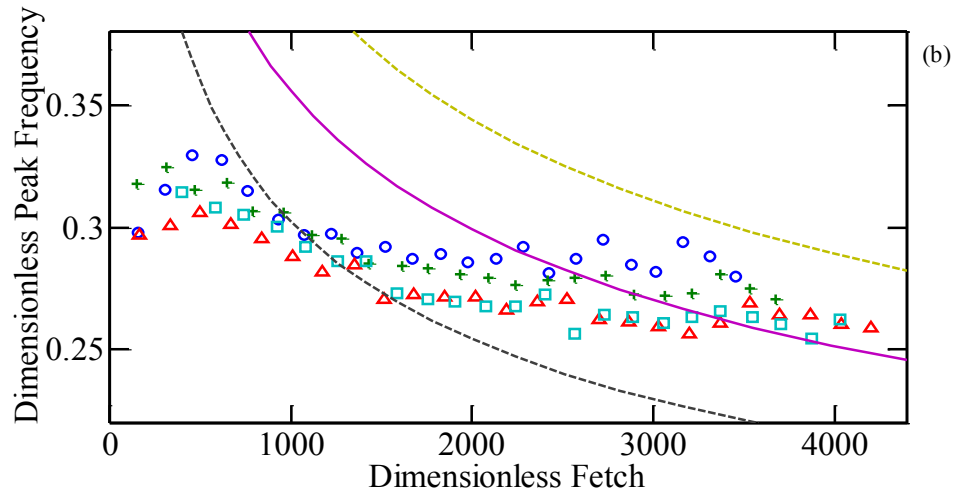

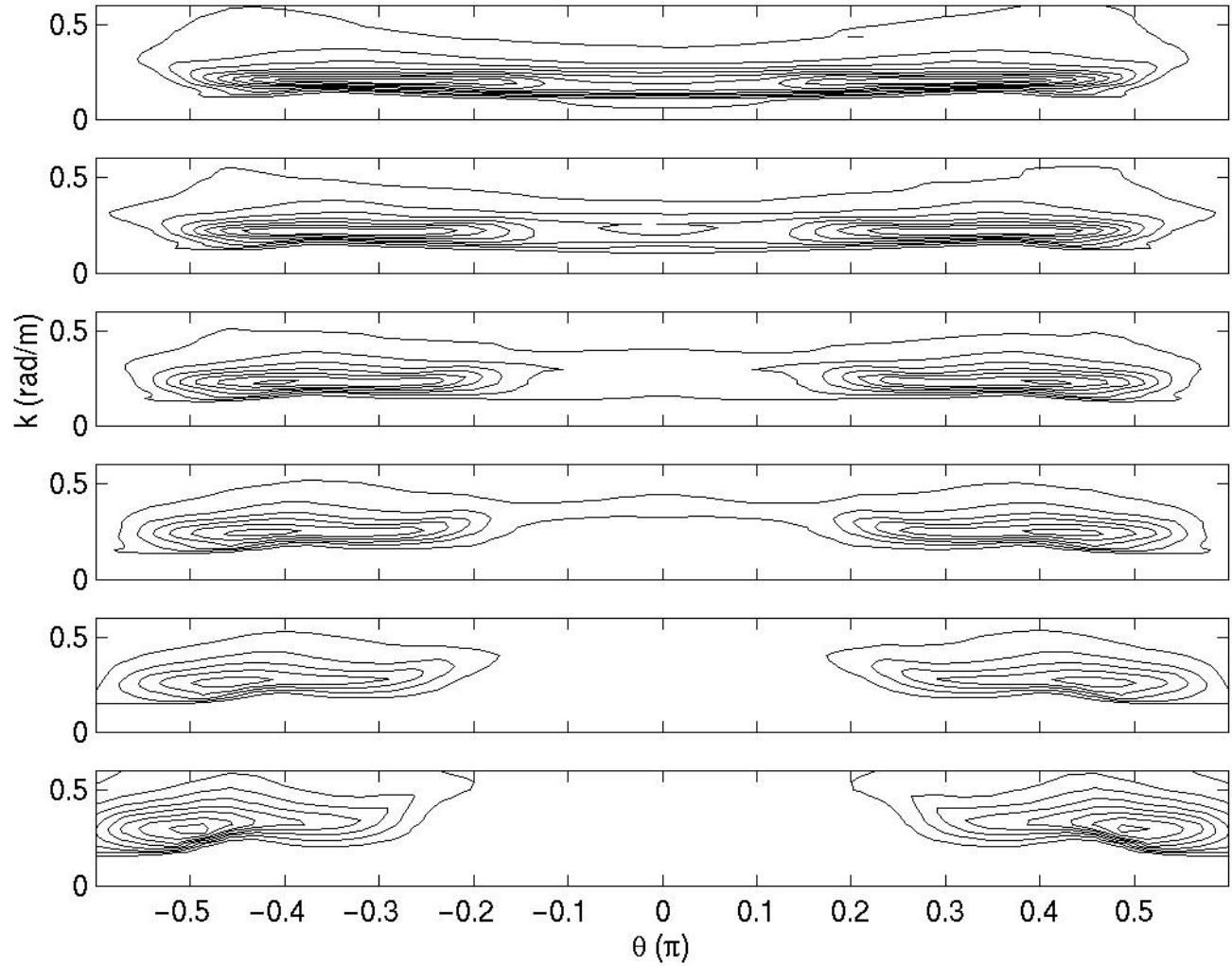

ATMGoM98Bimodal 14